\documentclass[journal]{IEEEtran}
\usepackage{cite}

\ifCLASSINFOpdf
 \usepackage[pdftex]{graphicx}
 \usepackage{makecell}
 \usepackage{mathtools}
 \usepackage[dvipsnames]{xcolor}
 \usepackage{multirow}
 \usepackage{caption}
 \usepackage{subcaption}
 \usepackage{diagbox}
 \usepackage{soul}
 \usepackage{cuted}
\else
\fi
\usepackage{amsmath}
\usepackage{array}
\begin{document}
%
\title{Hierarchical Frequency Control of Hybrid Power Plants Using Frequency Response Observer}
%
%
%

\author{Qian~Long,~\IEEEmembership{Member,~IEEE,}
        Kaushik~Das,~\IEEEmembership{Senior Member,~IEEE,}
        and~Poul~Sørensen,~\IEEEmembership{Fellow,~IEEE}
}

\maketitle

\begin{abstract}
Frequency control (FC) enables utility-scale grid-connected hybrid power plants (HPPs) to operate in compliance with grid code requirements while to capture value streams from provision of frequency control services (FCSs). In this paper, a novel hierarchical FC approach is proposed to allow HPPs to provide three types of FCSs, namely fast frequency response (FFR), frequency containment response (FCR) and frequency restoration response (FRR). To accommodate state-of-the-art fast FC, controllers for fast FCSs, such as FFR and FCR, are implemented at asset controllers, while controllers for slow FCSs like FRR are implemented at plant controllers or the HPP controller (HPPC). Control counteraction issue, which arises across control hierarchy, is then discussed. To solve this issue, an innovative frequency response observer (FROB) is proposed. Inspired by the concept of disturbance observer (DOB), FROB at plant controllers and the HPPC accurately estimates frequency response initiated at asset controllers, and the obtained estimation is used for control compensation at plant controllers and the HPPC to avoid control counteraction. This scheme achieves robust performance even when there are system uncertainties existing in HPPs, such as parameter uncertainty, unknown control malfunction, and time-varying communication delays. The proposed approach is implemented in a power system dynamic model in MATLAB/Simulink to highlight its effectiveness and robustness.
\end{abstract}


%

\section{Introduction}
%
%
%
%

\IEEEPARstart{A}{s} frequency stability has been challenged by the increasing share of inverter-based resources in modern power systems, the requirement for frequency control (FC) becomes demanding, posing challenges for the existing FC methods for power plants \cite{Eriksson2019}. Utility-scale hybrid power plants (HPPs) have received global attention due to enhanced controllability and efficient utilization of electrical infrastructure \cite{Gorman2020}. FC methodologies that enable HPPs to provide enhanced frequency response need to be further investigated, especially since there are additional number of assets and controllers in an HPP as compared to single-technology power plants \cite{Das2019}. However, there is little literature addressing FC of HPPs \cite{Pombo2019, Long2021FastSupercapacitors}.
\begin{figure}[t]
\centering
\includegraphics[scale=0.85]{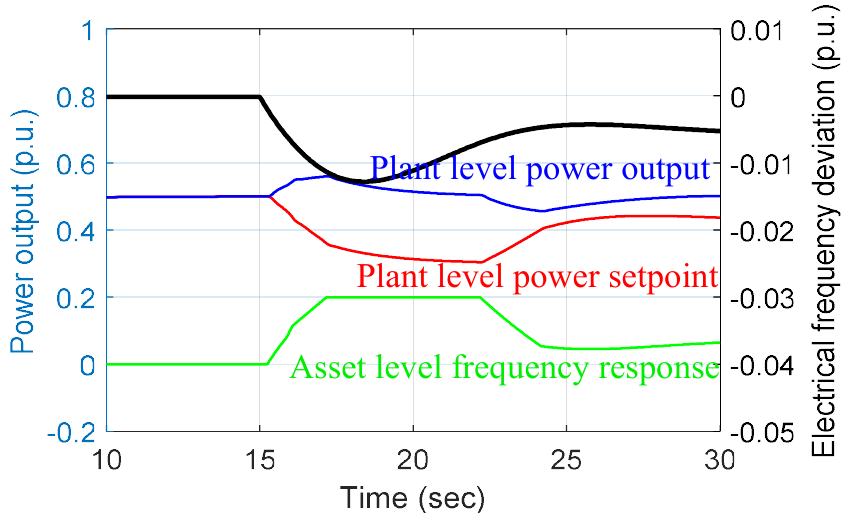}
\caption{An example of control counteraction during an under-frequency event.}
\label{fig:controlcounter}
\end{figure}

Hierarchical FC for single-technology power plants, which consists of asset controllers and a plant control, has been well established. A conventional approach is to introduce auxiliary FC loop for active power control at the plant controller, which generates time-varying active power setpoints for asset controllers when FC is activated \cite{Hansen2006, Karthikeya2014, Varma2020a, Calero2021}. While this approach enables power plants to provide slow frequency control services (FCSs) such as FRR, factors such as communication delays and limited control bandwidth could be the bottleneck for plant controllers to provide fast FCSs. To date, fast FC have been mostly implemented at asset controllers for wind turbines (WTs), photovoltaics (PVs) and energy storage (ESs) \cite{Nanou2015, Xin2013a, Hoke2017, Wu2018a, Licari2013, Liu2015, Wang2014}. Nevertheless, without any coordination strategy, when frequency response is activated at asset controllers, against which the plant controller will start to counteract, during a frequency event. Fig. \ref{fig:controlcounter} shows an example of such control counteraction. Therefore, a new hierarchical FC approach that supports fast FC at asset controllers and coordinates between asset controllers and plant controllers during the period of FCS provision is in need. Considering the HPP controller (HPPC) as an extra control level that provides control and supervision for wind power plants (WPPs), solar power plants (SPPs) and energy storage systems (ESSs), the design of the coordination strategy across the whole control hierarchy becomes even more complicated.

The main challenge of the coordination arises from frequency response being an unknown control action (as seen by the HPPC and plant controllers) from asset controllers. A possible way of addressing control counteraction could be estimating the total frequency response and compensating as the feedback at plant controllers and the HPPC. However, uncertainties existing in HPPs bring extra challenges into accurate estimation of frequency response. This leads to one of the innovations in the paper, which is to develop an observer that can accurately estimate the total frequency response activated at asset controllers given uncertainties in HPPs, for the purpose of avoiding counteraction across control hierarchy.

Disturbance observers (DOBs) algorithms for estimating system uncertainties and disturbances, have been used extensively in various power-related industrial applications. Time-domain DOBs, such as extended-state observer, have been utilized in the field of load frequency control to estimate total disturbance from loads and renewable generation \cite{Khousa2013, Liu2016}. In \cite{Chang2015}, similar DOB techniques have been applied to charging and discharging control of a flywheel ESS. Another category of DOBs, that is, frequency-domain DOBs, have been proposed and adopted in applications like load frequency control \cite{Saxena2013}, converter control \cite{Elkayam2019} and PV generation \cite{Sitbon2015}. The above-mentioned DOBs help generate a robust control system that achieves good performance in disturbance rejection and handles system uncertainties at the same time. This paper focuses on frequency-domain DOBs originally proposed in \cite{Ohishi1987} due to its simple and straightforward nature. However, it should be emphasized that frequency-domain DOBs can not be directly applied as a solution to the counteraction issue in hierarchical FC. It is because the design purpose of frequency-domain DOBs is to reject external disturbance, while frequency response from asset controllers is supposed to be accepted rather than rejected by plant controllers and the HPPC. Therefore, a new observer called frequency response observer (FROB), following the idea of frequency-domain DOBs, estimates the total frequency response from asset controllers as "the disturbance", and modification is introduced such that the closed-loop characteristic of FROB accepts frequency response. This modification also leads to the need to establish new design guidelines.

In this paper, a novel hierarchical FC approach using FROB is proposed to enable HPPs to provide coordinated FCSs. The main contribution of this approach is threefold: i) the proposed approach allows coordinated participation in frequency response from multiple technology power plants, such as WPPs, SPPs and ESSs; ii) the new control architecture allows HPPs to provide three types of FCSs including FFR, FCR and FRR, where controllers for fast FCSs like FFR and FCR are implemented at asset controllers and controllers for slow FCS like FRR is implemented at plant controllers and the HPPC. Under this architecture, state-of-the-art fast FC that is mainly developed at asset level is easy to apply; iii) an innovative FROB is introduced at plant controllers and the HPPC to avoid control counteraction against asset controllers. The proposed FROB achieves robust performance on estimating frequency response even when there are system uncertainties in HPPs.

Next section introduces dynamic modeling of HPPs with the proposed hierarchical FC approach. Section III starts with a brief introduction of the DOB and then focuses on FROB design and analysis. In Section IV, simulations are conducted and the effectiveness of the proposed control is illustrated. Section V concludes the paper.

\section{Dynamic Modeling and Control of HPP}

\subsection{HPP Component Model}
\begin{figure}[t]
\centering
\includegraphics[scale=0.85]{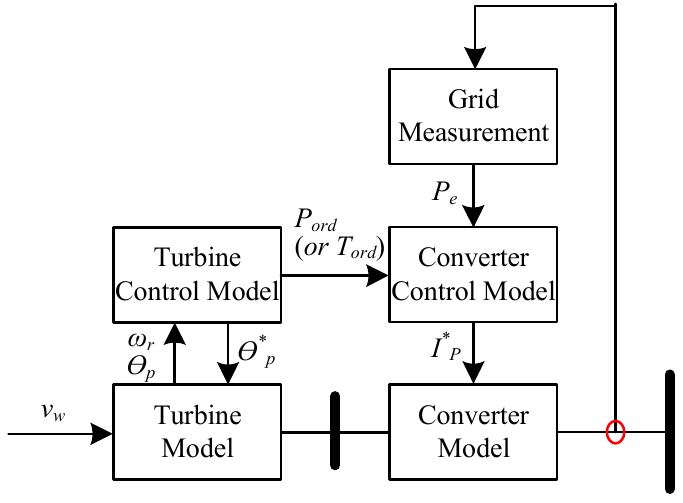}
\caption{WT dynamic model.}
\label{fig:wtmodel}
\end{figure}
\begin{figure}[t]
\centering
\includegraphics[scale=0.85]{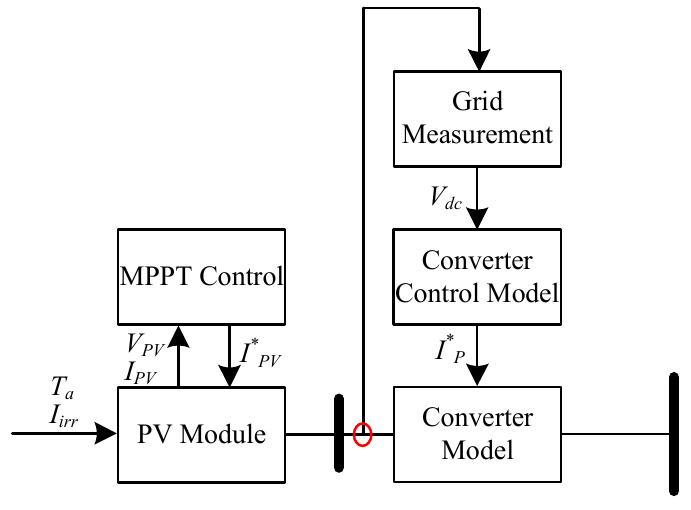}
\caption{PV dynamic model.}
\label{fig:pvmodel}
\end{figure}
\begin{figure}[t]
\centering
\includegraphics[scale=0.85]{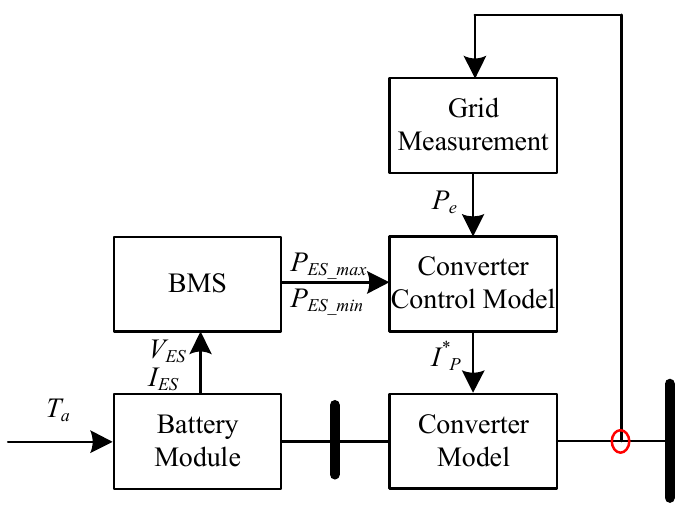}
\caption{ES dynamic model.}
\label{fig:esmodel}
\end{figure}

The dynamic models for WTs, PVs and ESs are shown in Fig. \ref{fig:wtmodel}, Fig. \ref{fig:pvmodel} and Fig. \ref{fig:esmodel}, respectively, with converter models and converter control models being developed. Since this paper is only concerned with FC, only active power control module is considered in converter control. Reactive power control module is beyond the scope. Modeling details of WTs, PVs and ESs for frequency studies can be found in \cite{Das2016, Margaris2012, Paduani2022ACapabilities, Calero2021}.

The WT dynamic model is built based on a variable speed Type 4 WT configuration, where the aerodynamic rotor of WT is directly coupled to the permanent magnet synchronous generator through a gearless drive train as described in IEC 61400-27-1 \cite{IEC2015}. For turbine model, aerodynamic model is included and a two-mass model is used for representing the mechanical shaft system of the WT. Turbine control model consists of two parts, torque control and pitch angle control. In order to control generator rotor speed, turbine control sends either power (or torque) command to converter control or pitch angle command to pitch actuator. Active power control module receives power (or torque) command from turbine control as the input, and generates active current command for WT converter model.

The PV dynamic model is built based on a generic WECC model for large-scale SPPs \cite{WECC2019}. The PV module takes weather conditions, such as ambient temperature and solar irradiance, as the input. Maximum power point tracking (MPPT) controller seeks for maximum power point of the PV module given weather conditions, PV terminal voltage and current. Active power control module receives DC voltage measurement, which is regulated by converter control, and generates active current command for PV converter model.

The ES dynamic model is built based on a generic WECC ES model \cite{WECC2016}. The battery module takes ambient temperature as the input. The battery management system (BMS) has been developed in order to estimate state of charge and to determine maximum discharging and charging current for ES converter control model. The input of the BMS includes ES terminal voltage and current. Active power control module takes external power command as the input, and generates active current command for ES converter model.

\subsection{Hierarchical Frequency Control}
\begin{figure*}[t]
\centering
	\includegraphics[scale=0.85]{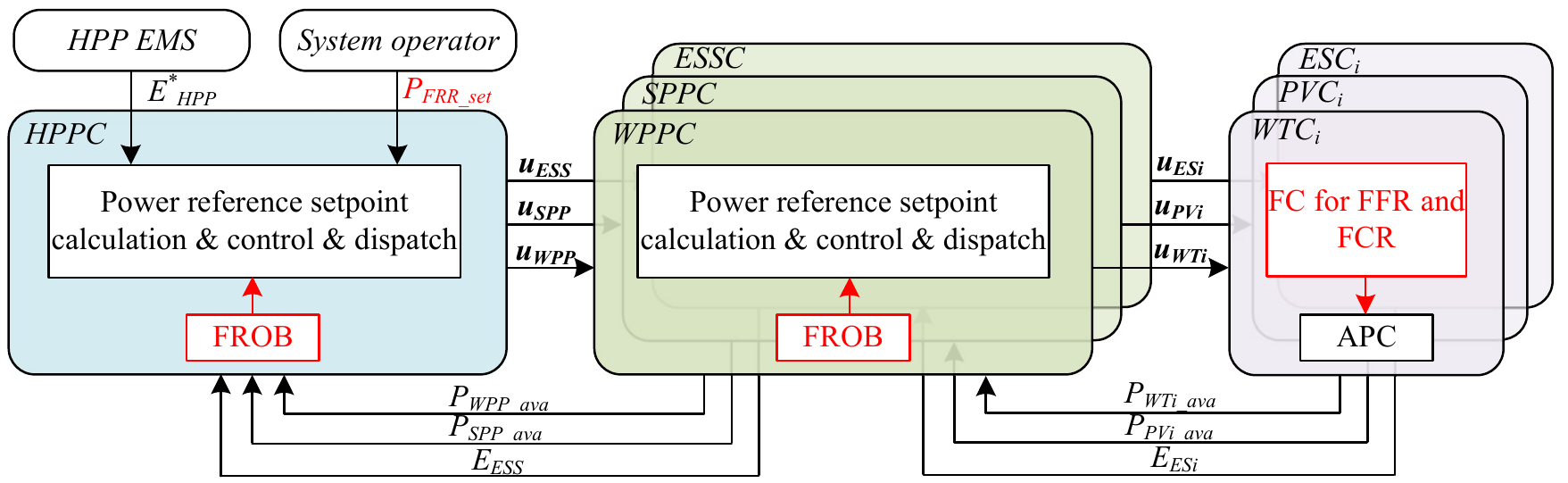}
 \caption{The proposed hierarchical FC architecture.}
 \label{fig:hierarchyfc}
\end{figure*}

In the existing approaches, controllers for FFR and FCR are integrated into plant controllers by adding auxiliary FC loops to active power control \cite{Pombo2019, Varma2020a}. However, these approaches could lead to a degraded frequency response due to several factors, such as communication delays or failures, control bandwidth and ramp rate limits. The above-mentioned factors restrict the speed of frequency response. The proposed hierarchical FC architecture is shown in Fig. \ref{fig:hierarchyfc}, consisting of three control levels, asset controllers, plant controllers and the HPPC. The main innovation of the proposed approach is that controllers for FFR and FCR are implemented at asset controllers. This approach allows fast FCSs, such as FFR and FCR, to be triggered at asset controllers by local frequency measurements. Therefore, no communication is involved during FCS provision. Meanwhile, control bandwidth of asset controllers is large enough to support fast FC, and control signal for fast FC is unaffected by ramp rate limits. FROB is implemented at both plant controllers and the HPPC to avoid control counteraction with asset controllers.

The references sent from the HPPC to plant controllers are defined as follows:
\begin{align} \label{eq:varhppc2wppc}
\begin{split}
\boldsymbol{u_{WPP}} = & [P^*_{WPP}, \boldsymbol{P^*_{WPP+}}, \boldsymbol{P^*_{WPP-}}, \boldsymbol{T_{FFR}}, \\ & DB_{FFR}, DB_{FCR}, R_{FCR}] \\
\end{split}
\end{align}

\begin{align} \label{eq:varhppc2sppc}
\begin{split}
\boldsymbol{u_{SPP}} = & [P^*_{SPP}, \boldsymbol{P^*_{SPP+}}, \boldsymbol{P^*_{SPP-}}, \boldsymbol{T_{FFR}}, \\ & DB_{FFR}, DB_{FCR}, R_{FCR}] \\
\end{split}
\end{align}

\begin{align} \label{eq:varhppc2essc}
\begin{split}
\boldsymbol{u_{ESS}} = & [P^*_{ESS}, \boldsymbol{P^*_{ESS+}}, \boldsymbol{P^*_{ESS-}}, \boldsymbol{T_{FFR}}, \\ & DB_{FFR}, DB_{FCR}, R_{FCR}] \\
\end{split}
\end{align}
where $P^*_{WPP}$, $P^*_{SPP}$, and $P^*_{ESS}$ are power references for WPP, SPP and ESS, respectively. $\mathbf{P^*_{WPP+}}$, $\mathbf{P^*_{WPP-}}$, $\mathbf{P^*_{SPP+}}$, $\mathbf{P^*_{SPP-}}$, $\mathbf{P^*_{ESS+}}$ and $\mathbf{P^*_{ESS-}}$ are upwards and downwards frequency reserve vectors for WPP, SPP and ESS, respectively. Each frequency reserve vector contains three elements representing the amount of FFR, FCR and FRR, respectively. $\mathbf{T_{FFR}}$ is a time vector that defines rising time, duration and falling time of FFR. $DB_{FFR}$, $DB_{FCR}$ and $R_{FCR}$ are frequency deadband of FFR, frequency deadband and droop coefficients of FCR, respectively. The references sent from plant controllers to asset controllers are defined as follows:
\begin{align} \label{eq:varwppc2wtc}
\begin{split}
\boldsymbol{u_{WTi}} = & [P^*_{WTi}, \boldsymbol{P^*_{WT+}}, \boldsymbol{P^*_{WT-}}, \boldsymbol{T_{FFR}}, \\ & DB_{FCR}, R_{FCR}] \\
\end{split}
\end{align}

\begin{align} \label{eq:varsppc2pvc}
\begin{split}
\boldsymbol{u_{PVi}} = & [P^*_{PVi}, \boldsymbol{P^*_{PV+}}, \boldsymbol{P^*_{PV-}}, \boldsymbol{T_{FFR}}, \\ & DB_{FCR}, R_{FCR}] \\
\end{split}
\end{align}

\begin{align} \label{eq:varessc2esc}
\begin{split}
\boldsymbol{u_{ESi}} = & [P^*_{ESi}, \boldsymbol{P^*_{ES+}}, \boldsymbol{P^*_{ES-}}, \boldsymbol{T_{FFR}}, \\ & DB_{FCR}, R_{FCR}] \\
\end{split}
\end{align}
where $P^*_{WTi}$, $P^*_{PVi}$, and $P^*_{ESi}$ are power references for the \textit{i}th WT, the \textit{i}th PV and the \textit{i}th ES, respectively. $\mathbf{P^*_{WT+}}$, $\mathbf{P^*_{WT-}}$, $\mathbf{P^*_{PV+}}$, $\mathbf{P^*_{PV-}}$, $\mathbf{P^*_{ES+}}$ and $\mathbf{P^*_{ES-}}$ are upwards and downwards frequency reserve vectors for the \textit{i}th WT, the \textit{i}th PV and the \textit{i}th ES, respectively. $\mathbf{T_{FFR}}$, $DB_{FFR}$, $DB_{FCR}$ and $R_{FCR}$ are the same FC settings as the ones passed from the HPPC. Note that these settings are rarely updated once fixed.
\begin{figure}[t]
\centering
	\includegraphics[scale=0.85]{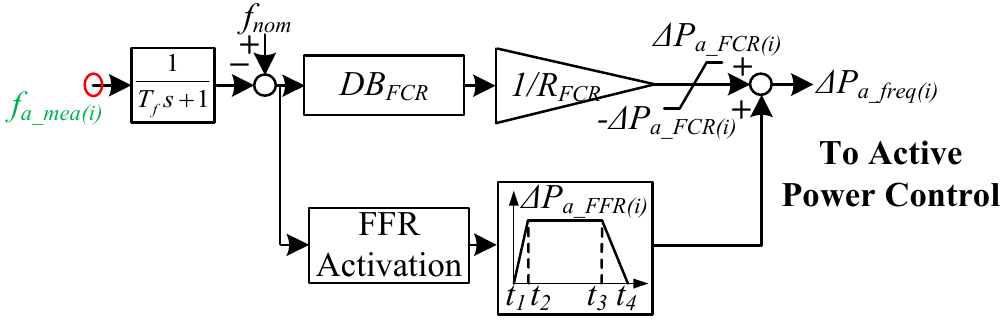}
 \caption{Asset controller design.}
 \label{fig:assetfc}
\end{figure}
\begin{figure}[t]
\centering
	\includegraphics[scale=0.85]{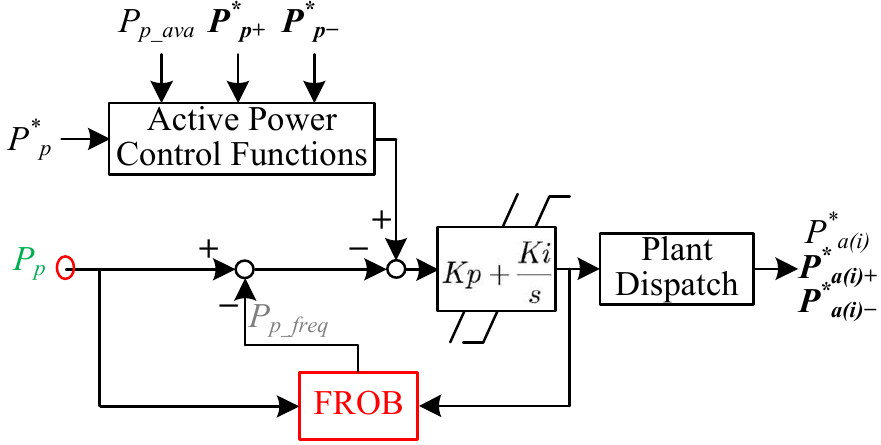}
 \caption{Plant controller design.}
 \label{fig:plantfc}
\end{figure}
\begin{figure}[t]
\centering
	\includegraphics[scale=0.85]{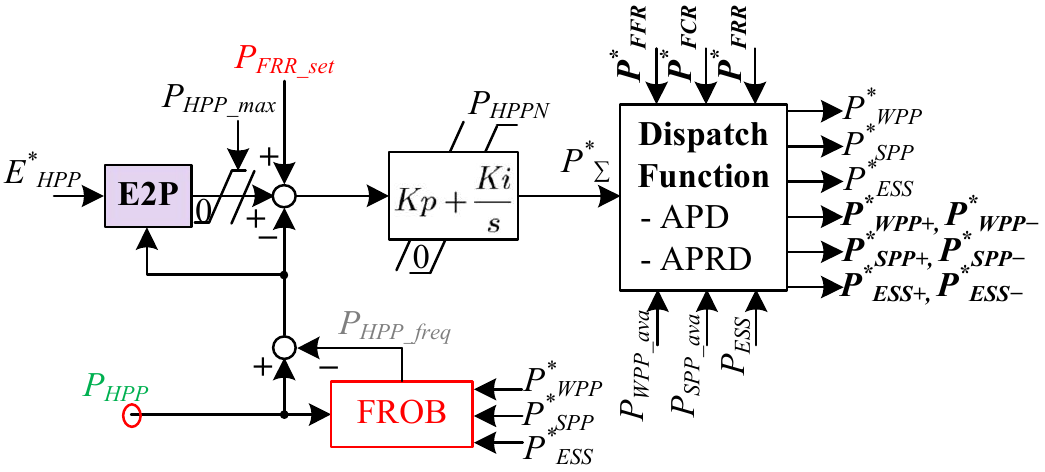}
 \caption{The HPPC design.}
 \label{fig:hppcfc}
\end{figure}

As shown in Fig. \ref{fig:assetfc}, a parallel control loop for FFR and FCR is implemented at asset controllers. Controller for FCR is implemented using a droop control with frequency deadband, while controller for FFR is implemented using an open-loop pre-defined shape with a FFR activation block. The output of this control loop is added to active power control of asset controllers \cite{Margaris2012}. It is worth pointing out that the proposed asset controller design is only an example and there are alternatives for implementing controllers for FFR and FCR with comparable performance \cite{Wu2018a}. To develop the optimal FC strategy from the assets is beyond the focus of the paper. Fig. \ref{fig:plantfc} shows the diagram of a plant controller, whose structure is universally applicable to the WPP controller, the SPP controller and the ESS controller. While no FC is implemented at the plant controller, a FROB is added to estimate the total frequency response from asset controllers. By subtracting the FROB output from active power measurement at plant point of connection (PoC), the compensated feedback only contains active power measurement irrelevant to FC. Another FROB is also added to the HPPC for the same purpose, as shown in Fig. \ref{fig:hppcfc}, allowing the HPPC to provide a stacked functionality of energy trade and FCS provision. Controller for FRR is implemented at the HPPC given that the response of FRR is slow, in the range of tens of minutes. An extra control command $P_{FRR\_set}$ is added to HPP active power controller to control the activation of FRR. The design details of plant controller and the HPPC could be found in the companion paper \cite{Long2021}.

\section{Frequency Response Observer}

The idea of FROB is inspired by frequency-domain DOBs which accurately track uncertainties and unknown disturbances. From the perspective of plant controllers and the HPPC, frequency responses activated at asset controllers can be considered as "unknown disturbances with uncertainties". The difference that distinguishes FROB from frequency-domain DOBs is that the total frequency response should be accepted rather than rejected at plant controllers and the HPPC. This section briefly reviews classical frequency-domain DOBs and then gives design and analysis for FROB.

\subsection{Overview of DOB}
\begin{figure}[t]
\centering
\includegraphics[scale=0.85]{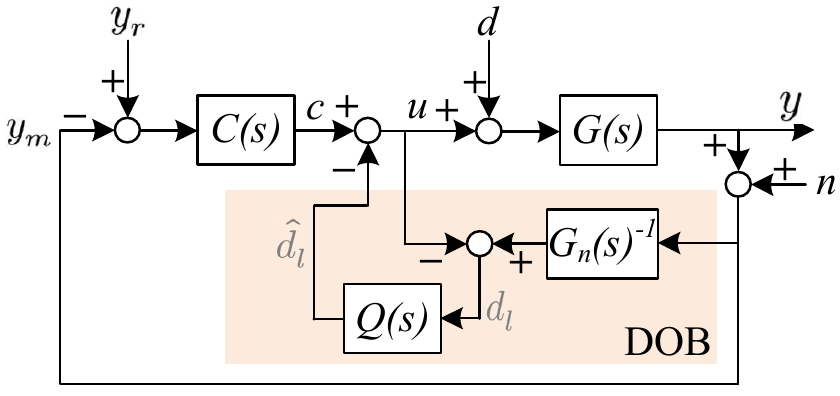}
\caption{A generic DOB structure.}
\label{fig:dob}
\end{figure}

Frequency-domain DOBs were initially proposed in \cite{Ohishi1987} to improve the robustness of the controllers for DC servo motor systems, given disturbances and uncertainties in the system. As shown in Fig. \ref{fig:dob}, the lumped disturbance $d_l(s)$ is written as \cite{Chen2016a}:
\begin{align} \label{eq:disturb}
\begin{split}
d_{l}(s) = & [G_n(s)^{-1} - G(s)^{-1}]y(s) + d(s) + G_n(s)^{-1}n(s) \\
\end{split}
\end{align}
where $d_l(s)$ captures all the disturbance and the uncertainties in the system. $G(s)$ and $G_n(s)$ are the physical system and the nominal model, respectively. $y(s)$ is the system output. $d(s)$ is the external disturbance. $n(s)$ is the measurement noise. The filter $Q(s)$ is used to estimate the lumped disturbance. The estimated lump disturbance $\hat{d}_{l}(s)$ is given by:
\begin{align} \label{eq:estdisturb}
\begin{split}
\hat{d}_{l}(s) = Q(s)d_{l}(s) \\
\end{split}
\end{align}
As the estimated lump disturbance is fed back to compensate the influence of the disturbance, the output of the equivalent system considering the DOB is expressed as:
\begin{align} \label{eq:equivalentsys}
\begin{split}
y(s) = G_{cy}(s)c(s) + G_{dy}(s)d(s) + G_{ny}(s)n(s)  \\
\end{split}
\end{align}
where $G_{cy}(s)$, $G_{dy}(s)$ and $G_{ny}(s)$ are transfer functions from $c(s)$, $d(s)$ and $n(s)$ to $y(s)$, respectively. In the frequency range where $Q(s) \approx 1$, (\ref{eq:equivalentsys}) reduces to \cite{Chen2016a}:
\begin{align} \label{eq:equivalentsys_re}
\begin{split}
y(j\omega) \approx G_{n}(j\omega)c(j\omega) - n(j\omega)  \\
\end{split}
\end{align}
The above representation implies the real physical system is forced to behave like the nominal system without disturbance given $Q(s) \approx 1$.

\subsection{FROB}
\begin{figure}[t]
  \centering
	\includegraphics[scale=0.85]{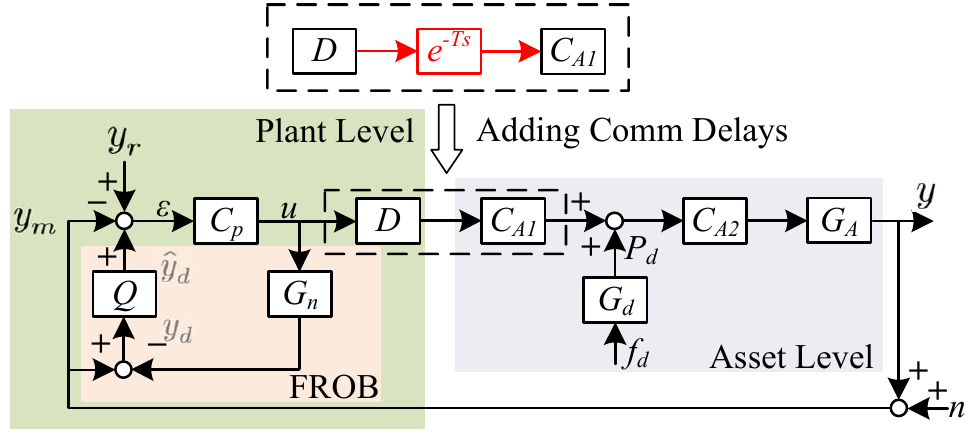}
	\caption{FROB design for plant controller.}
	\label{fig:frob}
\end{figure}

The main idea of FROB is to estimate frequency response provided by asset controllers, and then add the estimation to plant-level and HPP-level measurements as a compensation term, in order to avoid control counteraction. The FROB uses signals only from its own control level, and therefore communication is avoided during FCS provision. The FROB shown in Fig. \ref{fig:frob} is different from the original frequency-domain DOB design shown in Fig. \ref{fig:dob}. The DOB design adds the estimated disturbance directly to the output of the controller $C(s)$ to suppress disturbance and uncertainty. Here in the proposed FROB design, the estimated disturbance is added to the input of the controller $C_p(s)$ in order to compensate the influence of frequency response.

Fig. \ref{fig:frob} shows the diagram of FROB for plant controllers. The FROB diagram is generic to any type of power plant like WPP, SPP and ESS. Asset level includes controller before FC $C_{A1}$, controller after FC $C_{A2}$, asset dynamics $G_{A}$ and asset FC $G_{d}$. The input of $G_{d}$ is the measured frequency deviation $f_{d}$ and the output of $G_{d}$ is the FC command $P_{d}$. Output at asset level is the total power output $y$, and $n$ represents measurement noise. Plant level includes plant controller $C_p$, dispatch function $D$, and the FROB that consists of the nominal model $G_n$, and the FROB filter $Q$. The model $G_n$ is the nominal representation of $G$, where $G = DC_{A1}C_{A2}G_{A}$. The combined frequency response of FFR and FCR is estimated by taking the difference between $y_m$ and $G_nu$. Ideally, it is desired that
\begin{align}\label{eq:yd_est}
\begin{split}
\hat{y}_{d} & = Q(y_m - G_nu) \approx y_{d} \approx P_{d} C_{A2} G_{A} \\
\end{split}
\end{align}

\begin{figure}[t]
  \centering
	\includegraphics[scale=0.85]{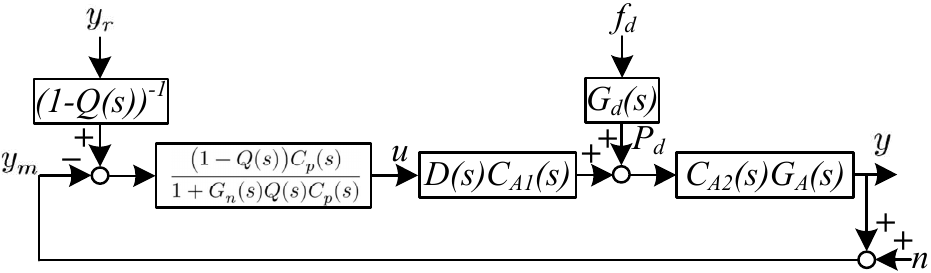}
	\caption{Linearized unity feedback system.}
	\label{fig:frob_linear}
\end{figure}

Based on Fig. \ref{fig:frob}, plant controller with FROB is restructured as a linearized unity feedback system, shown in Fig. \ref{fig:frob_linear}. Plant controller with FROB is equivalent to controller $C_p(s)$ reshaped by $Q(s)$ and $G_n(s)$, with a prefilter $\frac {1}{1-Q(s)}$ added to control reference $y_r$. The transfer function from control reference to output $G_{y_ry}(s)$ is defined as (\ref{eq:Gyry}). The transfer function from FC command to output $G_{P_dy}(s)$ is defined as (\ref{eq:Gpdy}). The transfer function from measurement noise to output $G_{ny}(s)$ is defined as (\ref{eq:Gny}). When the filter $Q(s) \approx 1$, the transfer functions are simplified into:
\begin{table*}
\centering
\begin{minipage}{0.75\textwidth}
\begin{align}
G_{y_ry}(s) & = \frac{C_p(s)G(s)}{1 + Q(s)C_p(s)G_n(s) - Q(s)C_p(s)G(s) + C_p(s)G(s)}
\label{eq:Gyry}
\end{align}
\begin{align}
G_{P_dy}(s) & = C_{A2}(s)G_A(s)\frac{1 + Q(s) C_p(s) G_n(s)}{1 + Q(s)C_p(s)G_n(s) - Q(s)C_p(s)G(s) + C_p(s)G(s)}
\label{eq:Gpdy}
\end{align}
\begin{align}
G_{ny}(s) & = -\frac{(1 - Q(s)) C_p(s) G(s)}{1 + Q(s)C_p(s)G_n(s) - Q(s)C_p(s)G(s) + C_p(s)G(s)}
\label{eq:Gny}
\end{align}
\medskip
\hrule
\end{minipage}
\end{table*}
\begin{align}
G_{y_ry}|_{Q \approx 1} & \approx \frac{C_pG}{1 + C_pG_n}
\label{eq:Gyrys}
\end{align}
\begin{align}
G_{P_dy}|_{Q \approx 1} & \approx C_{A2}G_A
\label{eq:Gpdys}
\end{align}
\begin{align}
G_{ny}|_{Q \approx 1} & \approx 0
\label{eq:Gnys}
\end{align}

It is seen from (\ref{eq:Gpdys}) and (\ref{eq:Gnys}) that, when $Q(s)$ is close to 1, frequency response is fully accepted in the closed-loop system, and measurement noise is completely attenuated. Although the unity might seem a perfect design for $Q(s)$, it would lead to zero loop transfer function, and modified characteristics of reference command tracking, seen from Fig. \ref{fig:frob_linear} and (\ref{eq:Gyrys}). Only when the nominal model is correct ($G_n=G$) does $G_{y_ry}(s)$ stay unchanged, and plant controller $C_p$ and the FROB filter can be designed independently. Therefore, low-pass filters have been used for the design of $Q(s)$.
\begin{figure}[t]
  \centering
	\includegraphics[scale=0.85]{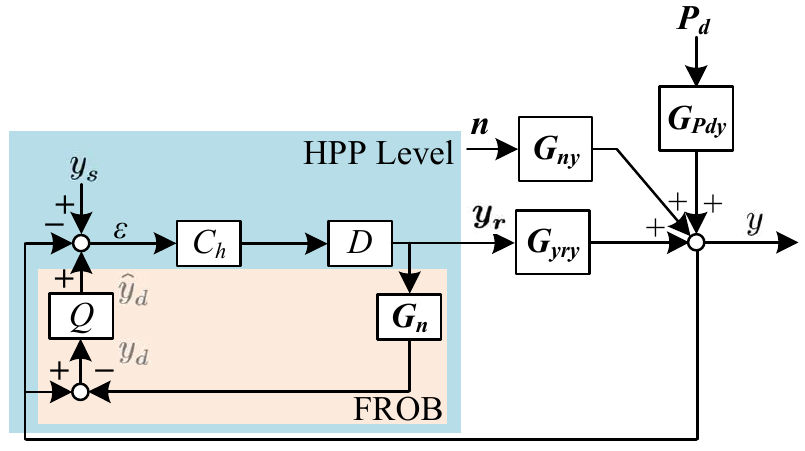}
	\caption{FROB design for HPPC.}
	\label{fig:frob_hppc}
\end{figure}

FROB for the HPPC is similar to FROB for plant controllers, as shown in Fig. \ref{fig:frob_hppc}. The difference lies mainly in the model being used in FROB to represent the nominal model of the HPP, $\mathbf{G_n}$, which is a multiple-input single-output model. The actual HPP dynamics are presented using the closed-loop transfer functions including both plant level and asset level. These closed-loop transfer functions for WPPs, SPPs and ESSs are obtained using (\ref{eq:Gyry}), (\ref{eq:Gpdy}) and (\ref{eq:Gny}). A similar linearized unity feedback system to Fig. \ref{fig:frob_linear} can be easily derived including HPP level, plant level and asset level. To avoid redundancy, only plant level and asset level are considered as the example in the following design guideline and robustness analysis. Nevertheless, they can readily be applied to the system including all three levels.

\subsection{Design Guidelines}
\begin{table}[t]
    \begin{tabular}{m{8cm}}
    \hline
    \textbf{Algorithm 1 $Q(s)$ Selection} \\
    \hline
    \hangindent=1.7em 1: \hspace{0.6em}Initialize the algorithm by setting up a general structure for $Q(s)$: $Q(s) = \frac{1 + \sum_{m=1}^{n-1} f_m s^m}{1 + \sum_{m=1}^{n} f_m s^m}$ \\
    \hangindent=1.7em 2: \hspace{0.6em}Start the filter degree $n = 1$ and choose $\omega_c$ to be equal to or larger than the frequency of measurement noises. \\
    \hangindent=1.7em 3: \hspace{0.6em}Perform bode analysis of $G_{P_dy}$ and $G_{ny}$, and fine tune $\omega_c$, such that $G_{P_dy}$ has unity gain around the frequency range of frequency response and $G_{ny}$ has small gain around the frequency range of measurement noises. \\
    \hangindent=1.7em 4: \hspace{0.6em}Increase the filter degree by 1 and choose the butterworth filter coefficients while keeping the same $\omega_c$. Check if $G_{P_dy}$ has unity gain for a wider frequency range, that is, a preferable performance on frequency response acceptance, and $G_{ny}$ has a smaller gain for a wider frequency range, that is, a preferable performance on measurement noise attenuation. \\
    \hangindent=1.7em 5: \hspace{0.6em}\textbf{If} $G_{P_dy}$ and $G_{ny}$ show a better performance with the increased filter degree \\
    6: \hspace{2.2em} \textbf{Repeat} Step 4. \\
    \hangindent=1.7em 7: \hspace{0.6em}\textbf{Else} \\
    8: \hspace{2.2em} Finalize the design of $Q(s)$ with the original degree. \\
    \hline
    \end{tabular}
    \label{tab:qselect}
\end{table}

The design of the filter $Q(s)$ plays a significant role in the FROB. The main requirement is that $Q(s)$ should be a low-pass filter with unity gain for low frequency to accept frequency response and zero gain for high frequency to attenuate measurement noise. Unlike the DOB filter design \cite{Umeno1993RobustManipulators, Schrijver2002}, no degree requirements exist for the FROB filter since there is no inverse of the model. There is an amount of freedom to select $Q(s)$ as long as $Q(j\omega) \approx 1$ in the frequency range covering frequency response and measurement noise. The main parameters to be selected for $Q(s)$ are the cut-off frequency $\omega_c$ and the filter degree $n$. A straightforward algorithm is developed to select the filter $Q(s)$, shown as Algorithm 1.
\begin{figure}[t]
  \centering
	\includegraphics[scale=0.85]{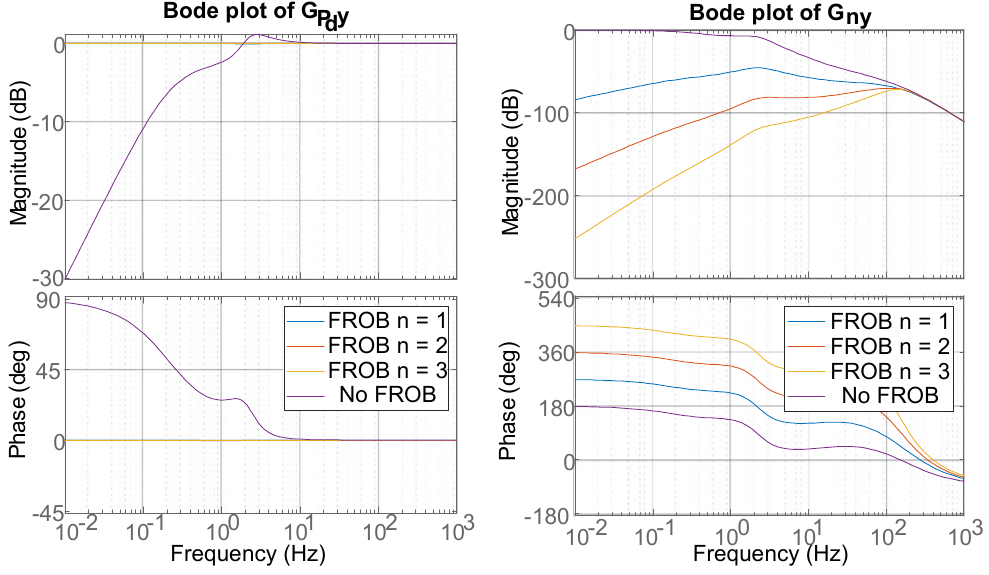}
	\caption{Bode plots of transfer functions - $G_{P_dy}$ and $G_{ny}$.}
	\label{fig:bodeplot_Gpdy_Gny}
\end{figure}

Fig. \ref{fig:bodeplot_Gpdy_Gny} presents bode plots of $G_{P_dy}(s)$ and $G_{ny}(s)$. Without FROB being included, $G_{P_dy}(s)$ shows the rejection of frequency response in low frequency range, which explains the importance of the FROB in hierarchical FC. With the filter degree equal to 1, 2 and 3, $G_{P_dy}(s)$ shows the characteristics of frequency response acceptance. It is also seen that without FROB, $G_{ny}(s)$ shows a limited performance of measurement noise attenuation. With a higher filter degree, the performance of measurement noise attenuation has been improved. Therefore, the degree of 3 is preferable for the design of $Q(s)$.

PI gains of plant controllers are selected based on the consideration of coordination with other control levels, and performance requirements of command reference tracking. The bandwidth of the closed-loop system including plant level and asset level is selected to be 0.25 Hz and the phase margin is selected to be 150$^{\circ}$.

\subsection{Robustness Analysis}
In this subsection, robustness analysis under two scenarios are considered for the FROB: robustness in the presence of uncertain parameters and time-varying communication delays. When uncertainties are considered, it implies that $G_n(s) = G(s)$ is invalid.

In the first scenario, assume that asset dynamics has changed to a different system $G'_A$, leading to the actual physical system $G_{m1}$ different from $G_n$, which is written as:
\begin{align}
G_{m1} & = DC_{A1}C_{A2}G'_{A}
\label{eq:Gm1}
\end{align}
The communication delay exists between asset controllers and plant controllers. In the second scenario, a time-varying communication delay is introduced as $e^{-Ts}$ as shown in Fig. \ref{fig:frob}, and $T$ is uncertain but is known to lie in the range [$T_1$, $T_2$]. The integration of communication delays leads to another physical system $G_{m2}$, which is written as:
\begin{align}
G_{m2} & = e^{-Ts}G = e^{-Ts}DC_{A1}C_{A2}G_{A}
\label{eq:Gm2}
\end{align}
Based on the perturbation multiplicative model \cite{doyle1992}, the magnitude bound for system stability in these two scenarios is written as:
\begin{align}
M_1(s) & = \frac{G'_A(s)}{G_A(s)} -1 \label{eq:M1} \\
M_2(s) & = e^{-Ts} -1 \label{eq:M2}
\end{align}
The loop transfer function of the FROB including an asset controller and a plant controller is derived from Fig. \ref{fig:frob_linear}:
\begin{align}
L(s) & = \frac{(1-Q(s))C_p(s)G(s)}{1+Q(s)G_n(s)C_p(s)}
\label{eq:looptransfer}
\end{align}

To check the stability of the closed-loop system in the presence of uncertain parameters, it should be guaranteed that
\begin{align}
|M_1(j\omega)| & < |1 + \frac{1}{L(j\omega)}| \textrm{  for all }\omega \textrm{ and } a_{min} \leq a \leq a_{max}\label{eq:robuststability_M1} \\
|M_2(j\omega)| & < |1 + \frac{1}{L(j\omega)}|\textrm{  for all }\omega \textrm{ and } T_1 \leq T \leq T_2\label{eq:robuststability_M2}
\end{align}
Since $a$ and $T$ are uncertain, the magnitude of the LHS of (\ref{eq:robuststability_M1}) and (\ref{eq:robuststability_M2}) are not known. As long as functions $W_1(s)$ and $W_2(s)$ are found, such that
\begin{align}
|\frac{G'_A(j\omega)}{G_A(j\omega)} - 1| & < |W_1(j\omega)|\textrm{  for all }\omega \textrm{ and } a_{min} \leq a \leq a_{max}\label{eq:upperbound_W1}\\
|e^{-j\omega T} - 1| & < |W_2(j\omega)|\textrm{  for all }\omega \textrm{ and } T_1 \leq T \leq T_2\label{eq:upperbound_W2}
\end{align}
the robust stability conditions with uncertain parameters and time-varying communication delay are satisfied by
\begin{align}
|W_1(j\omega)| < |1 + \frac{1}{L(j\omega)}|\textrm{  for all }\omega \label{eq:robuststability2}\\
|W_2(j\omega)| < |1 + \frac{1}{L(j\omega)}|\textrm{  for all }\omega \label{eq:robuststability2}
\end{align}

\begin{figure}[t]
\centering
\begin{subfigure}{0.24\textwidth}
\includegraphics[width = \textwidth]{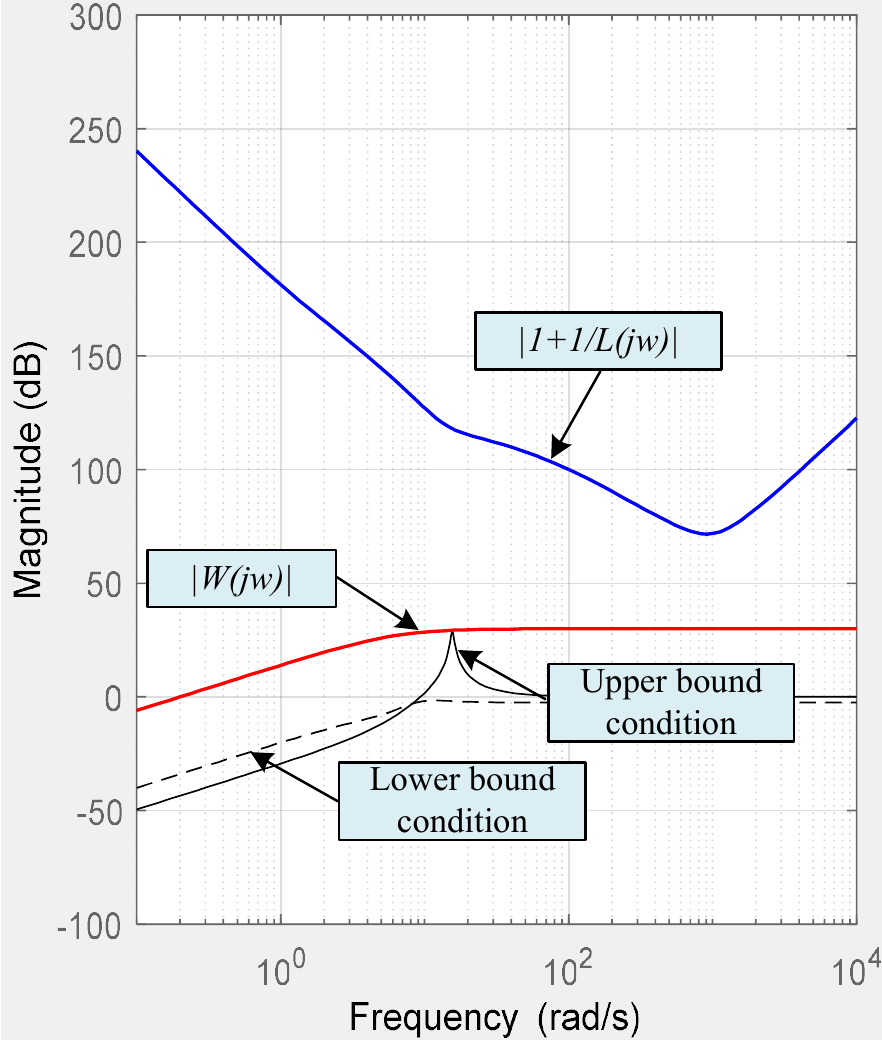}
\caption{$M_1(s) = \frac{G'_A}{G_A} - 1$}
\label{fig:magplotmismatch}
\end{subfigure}
\hfill
\begin{subfigure}{0.24\textwidth}
\includegraphics[width = \textwidth]{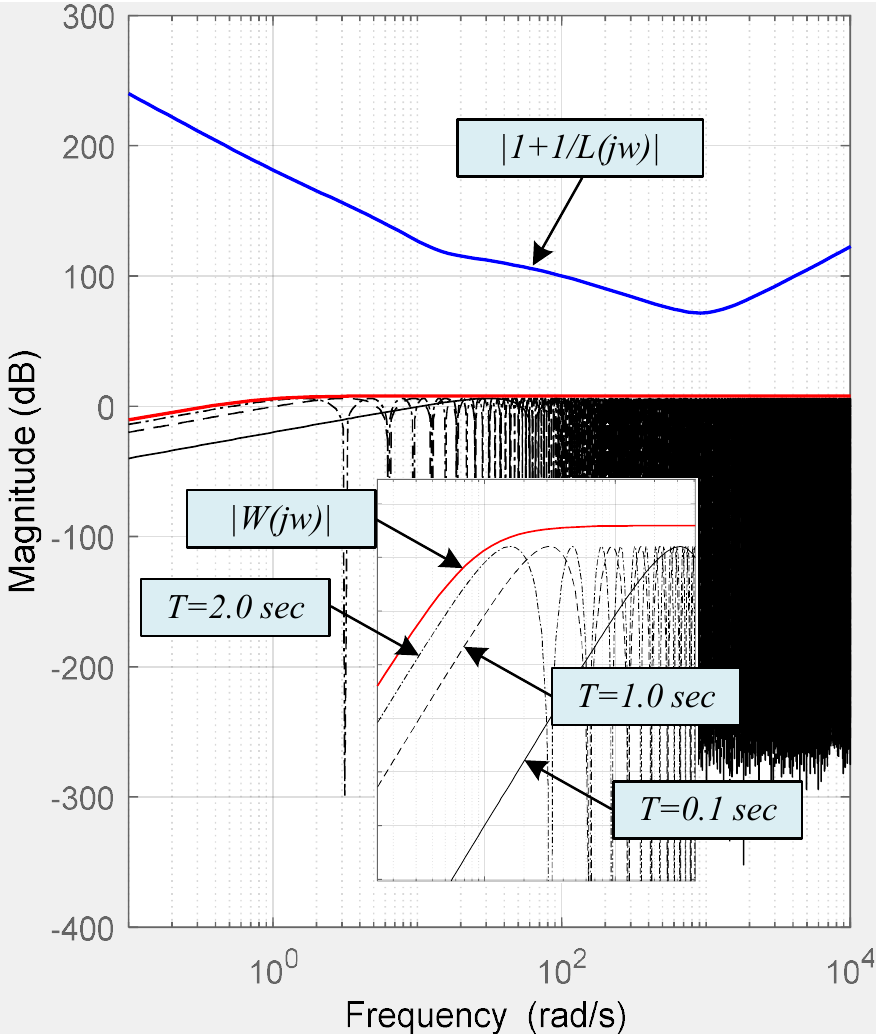}
\caption{$M_2(s) = e^{-Ts} - 1$}
\label{fig:magplotdelay}
\end{subfigure}
\caption{Magnitude plot of $|M(j\omega)|$, $|W(j\omega)|$ and $|1 + \frac{1}{L(j \omega)}|$.}
\label{fig:magplot}
\end{figure}

Fig. \ref{fig:magplot} shows magnitude plots for $|M(j\omega)|$, $|W(j\omega)|$ and $|1 + \frac{1}{L(j \omega)}|$ with two types of uncertainties. In Fig. \ref{fig:magplotmismatch}, it can be seen that, when the actual dynamics vary around the nominal model, a function $W_1(s)$ can be obtained as the upper bound of $|\frac{G'_A(j\omega)}{G_A(j\omega)} - 1|$. Likewise, based on industrial practice, communication delay between plant controllers and asset controllers ranges from 0 to 2 sec. The function $W_2(s)$ is obtained as the upper bound of $|e^{-j\omega T} - 1|$. It can be seen that the robust stability condition is met for both scenarios. Also a high magnitude of $|1 + \frac{1}{L(j \omega)}|$ indicates the FROB results in a very robust system.

\section{Simulation Results}

In order to validate the proposed hierarchical FC approach as well as the relevant analytical results, three case studies are carried out using a single-bus power system model with system parameters shown in \cite{Das2016}. This power system model, along with a detailed HPP model including the WPP, the SPP and the ESS and the corresponding controls, is built in MATLAB/Simulink\texttrademark. The first case study is the benchmark case showing the performance of the proposed FC on the coordination of three types of FCSs. The comparison with the centralized implementation is presented in the second case study. The focus of the third case study is on the robustness of the FROB when different types of uncertainties are included in the system.

\subsection{Benchmark Studies}
\begin{figure}[t]
  \centering
  \begin{subfigure}{0.45\textwidth}
  \includegraphics[width = \textwidth]{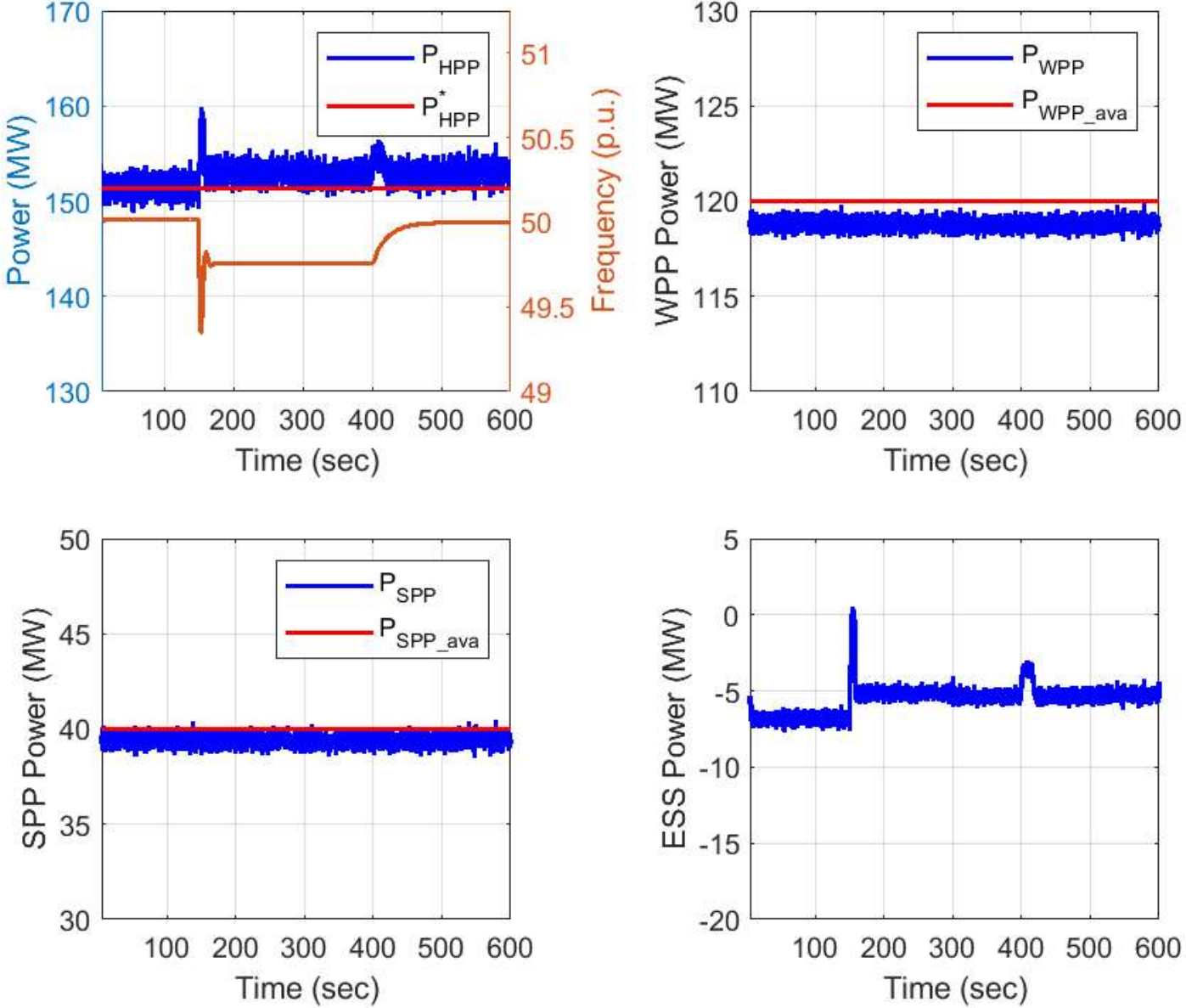}
  \caption{Only ESS participates in FCS provision with HPP power reference equal to 0.9 p.u. and HPP power reserve reference equal to 0.05 p.u.}
  \label{fig:benchmark1}
  \end{subfigure}
  \begin{subfigure}{0.45\textwidth}
  \includegraphics[width = \textwidth]{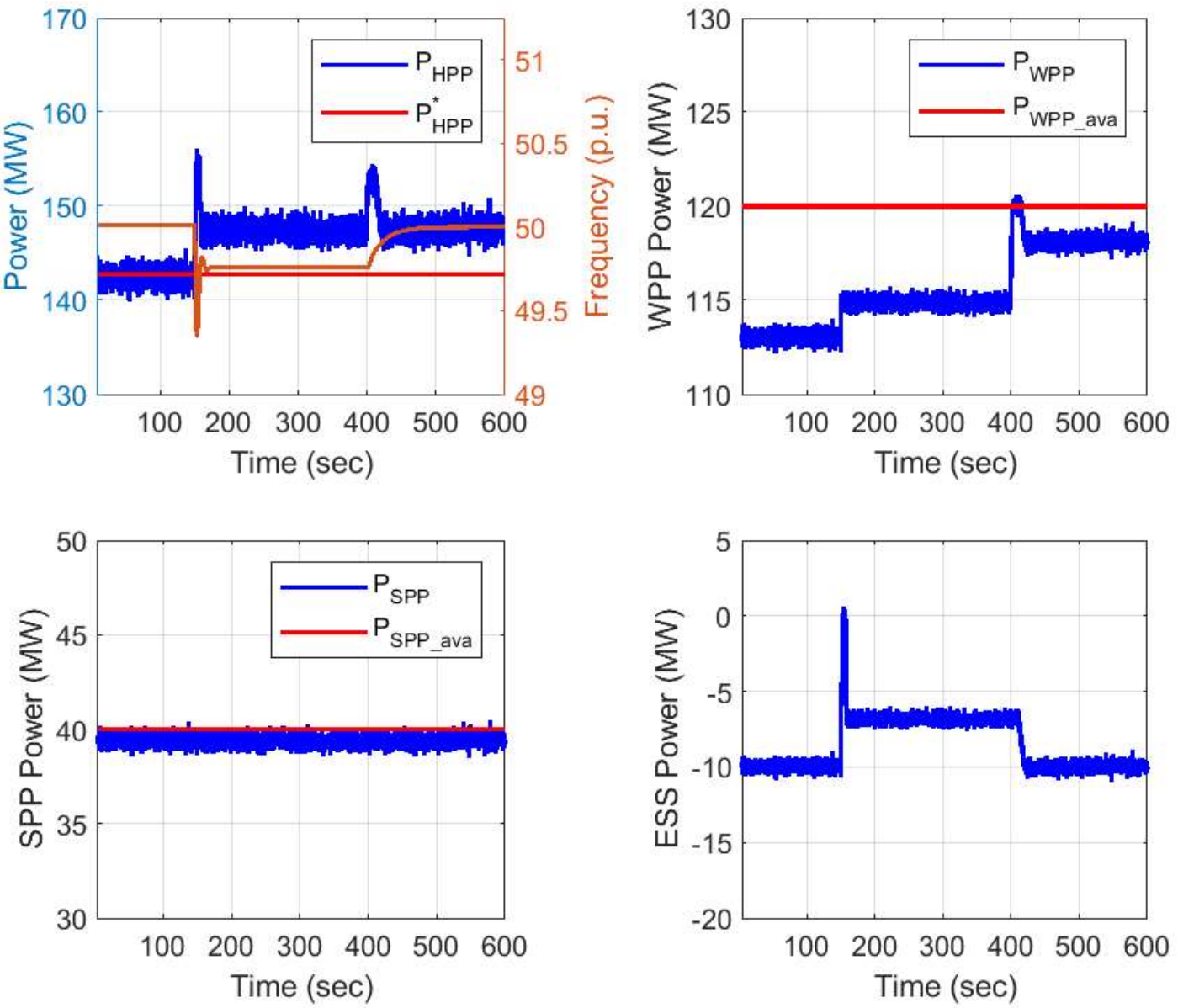}
  \caption{Both WPP and ESS participate in FCS provision with HPP power reference equal to 0.84 p.u. and HPP power reserve reference equal to 0.10 p.u.}
  \label{fig:benchmark2}
  \end{subfigure}
  \caption{FCS provision of an HPP during an under-frequency event.}
  \label{fig:benchmark}
\end{figure}

The under-frequency event is triggered at t = 150 sec by a heavy load switching-in within the power system. A secondary FC is initiated by system operator at t = 400 sec to bring system frequency back to nominal. Fig. \ref{fig:benchmark} shows frequency response provided by the HPP while the HPP tracks power setpoint decided by the HPP EMS. Fig. \ref{fig:benchmark1} presents the case when all the FCSs including FFR, FCR and FRR are fulfilled by the ESS. At t = 150 sec when the frequency dip occurs, the ESS provides a combined frequency response including FFR and FCR. When the system frequency reaches steady-state, only FCR is still activated and the charging power of the ESS achieves a new steady-state value. At t = 400 sec, the FRR setpoint is sent to the HPPC by system operator, with the frequency restored back to 50 Hz due to the effect of system-level secondary frequency response. Meanwhile, FCR provided by the HPP dies away when the frequency deviation reduces to zero.

Fig. \ref{fig:benchmark2} presents another case when FCSs are fulfilled by both the ESS and the WPP. According to the active power reserve dispatch algorithm in \cite{Long2021}, the capacity of the ESS is reserved to provide FFR and FCR, and the WPP runs at the derated operation to provide FCR and FRR. When the frequency dip occurs at t = 150 sec, the ESS provides a combined frequency response including FFR and FCR while the WPP provides FCR. At t = 400 sec, the FCR dies away due to frequency restoration, so it can be seen that the charging power of the ESS returns to the pre-event value. Meanwhile, the WPP participates in secondary frequency response initiated by system operator via the contribution of FRR.

The above results show that the proposed hierarchical FC is effective in coordinating three types of FCSs, namely FFR, FCR and FRR, during frequency events. Besides, the approach works in harmony with conventional active power control. Thanks to FROB, no counteraction is identified. More thorough operating conditions will be considered to validate the proposed approach in the future work.

\subsection{Comparison with Centralized Approach}
\begin{figure}[t]
  \centering
  \includegraphics[scale = 0.85]{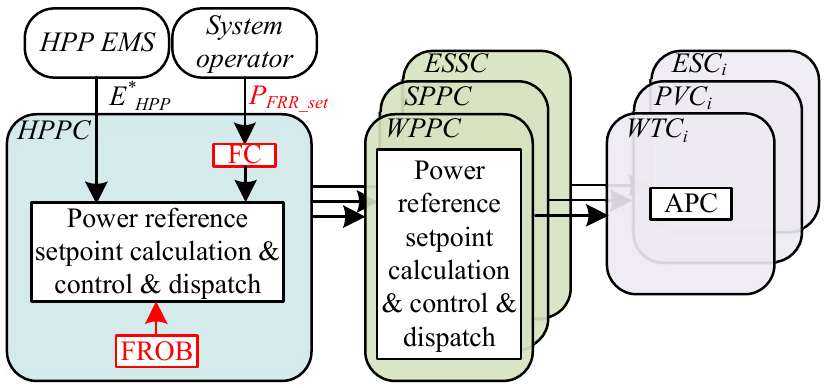}
  \caption{Centralized implementation of hierarchical FC.}
  \label{fig:centralizedfc}
\end{figure}
\begin{figure}[t]
  \begin{subfigure}{0.24\textwidth}
  \includegraphics[width = \textwidth]{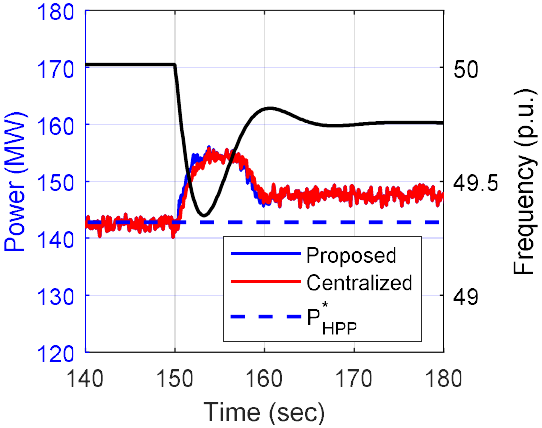}
  \caption{$T_{cd} = 0.0 s$.}
  \label{fig:comparison1}
  \end{subfigure}
  \hfill
  \begin{subfigure}{0.24\textwidth}
  \includegraphics[width = \textwidth]{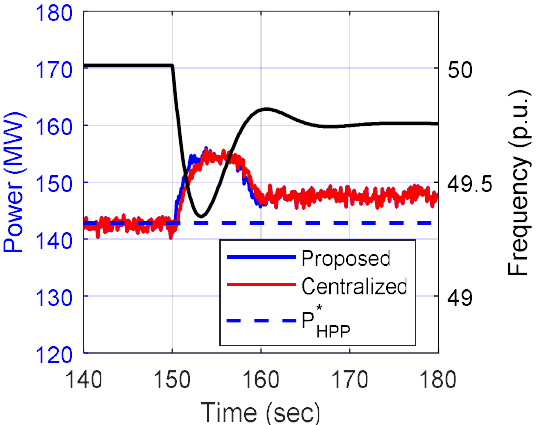}
  \caption{$T_{cd} = 0.1 s$.}
  \label{fig:comparison2}
  \end{subfigure}
  \medskip
  \begin{subfigure}{0.24\textwidth}
  \includegraphics[width = \textwidth]{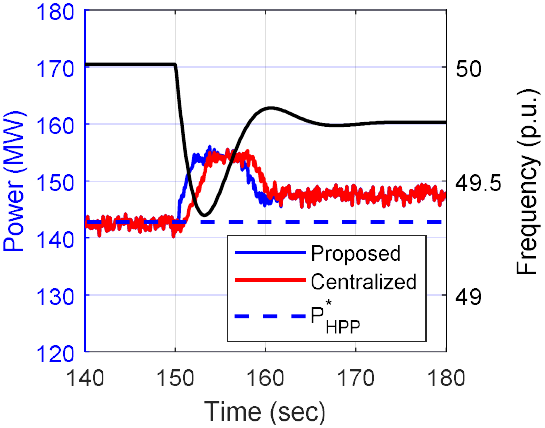}
  \caption{$T_{cd} = 1.0 s$.}
  \label{fig:comparison3}
  \end{subfigure}
  \hfill
  \begin{subfigure}{0.24\textwidth}
  \includegraphics[width = \textwidth]{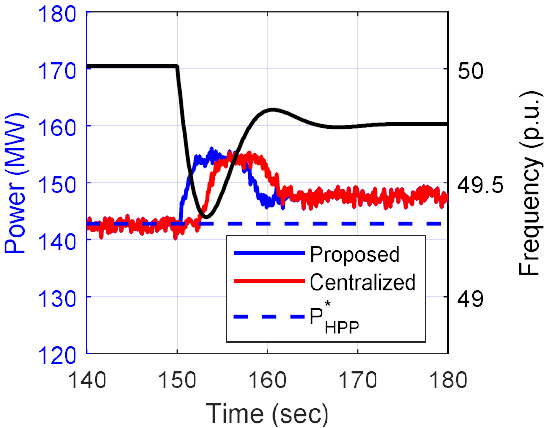}
  \caption{$T_{cd} = 2.0 s$.}
  \label{fig:comparison4}
  \end{subfigure}
  \caption{Performance comparison of FCS provision between the proposed approach and the centralized approach considering communication delays.}
  \label{fig:comparison}
\end{figure}

Instead of at each asset, FC can be implemented at the HPPC \cite{Pombo2019}. A simplified diagram of architecture design with centralized implementation of FC is shown in Fig. \ref{fig:centralizedfc}. The main difference between the proposed approach and the centralized implementation is that the controller for FFR and the FCR is located at the HPPC and activated by the frequency measured at the HPP PoC. The active power setpoints obtained from the HPPC are passed and added to the active power control at each plant controller. In this case, FROB is still necessary at the HPPC to distinguish frequency response from active power tracking, but is unnecessary at plant controllers.
\begin{table}[t]
\centering
    \caption{Impact of Communication Delay on Frequency Response Provision }
    \label{tab:cdonfrp}
 \resizebox{.48\textwidth}{!}{
\begin{tabular}{ c | c | c }
 \hline
 \multirow{2}{*}{Comm Delay $T_{cd}$ (sec)} & \multicolumn{2}{c}{Rise Time (sec)} \\
 \cline{2-3} 
  & Distributed & Centralized \\
 \hline\hline
 0.0 & 1.9 & 2.9 \\
 0.1 & 1.9 & 2.9 \\
 1.0 & 1.9 & 3.4 \\
 2.0 & 1.9 & 4.6 \\
 \hline
\end{tabular}
}
\end{table}

Fig. \ref{fig:comparison} presents the performance of FFR and FCR provision considering that communication delay exists between asset controllers and plant controllers. The communication delay $T_{cd}$ is assumed to lie in the interval between 0 sec and 2 sec, and values including 0.0 sec, 0.1 sec, 1.0 sec and 2.0 sec are selected. This assumption is realistic for a large renewable power plant \cite{Shi2020Data-DrivenFarm}. It is obviously seen that with no or small communication delay, frequency response via the proposed approach and the centralized approach are comparable. However, for larger communication delays like 1 sec or 2 sec, the centralized approach provides a delayed frequency response that doesn't meet technical requirement of FFR and FCR \cite{Eriksson2019}. As shown in Table \ref{tab:cdonfrp}, even if there is negligible communication delay, time constant of frequency response via the centralized approach is still larger than the proposed approach. It is because for the centralized approach, how fast frequency response is activated is subject to control bandwidth of plant controller. On the contrary, the proposed approach provides the same level of frequency response regardless of the communication delay.

\subsection{Study on FROB Robustness}
In order to investigate the robustness of FROB, three types of uncertainties are considered in this subsection. The first type is uncertainty in system parameters. The second type is a random malfunction, which is unknown to the plant controller, of FC at asset level. The third type refers to time-varying communication delay between asset controllers and plant controllers. Either type of uncertainty leads to a nominal model $G_n(s)$ that doesn't represent the actual system $G(s)$. For simplicity purpose, only asset controllers and plant controllers are included in this study. However, the conclusion also applies to the case when the HPPC is considered.

To evaluate the performance of FROB, ideal responses of FFR and FCR are defined by the performance obtained from an open-loop plant controller. When the plant controller is open-loop, there is no counteraction between asset controllers and plant controllers during the frequency event. Another coordination strategy to avoid counteraction is to estimate the total frequency response from asset controllers using FC settings that are made available to plant controllers, plus the frequency measured at the plant PoC. In the following studies, the above-mentioned strategies are named Strategy \#1 and Strategy \#2, respectively. The proposed approach is named Strategy \#3.

\subsubsection{Impact of uncertain system parameters}
\begin{figure}[t]
  \centering
  \begin{subfigure}{0.24\textwidth}
  \includegraphics[width = \textwidth]{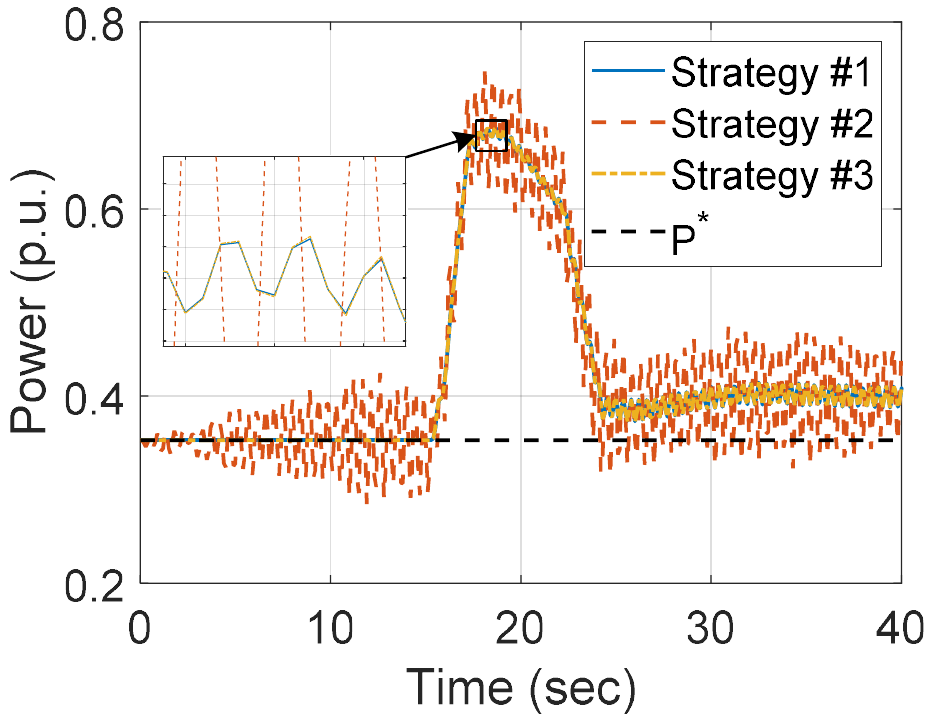}
  \caption{upper bound uncertainties}
  \label{fig:frob_robust_ucp_upper}
  \end{subfigure}
\hfill
  \begin{subfigure}{0.24\textwidth}
  \includegraphics[width = \textwidth]{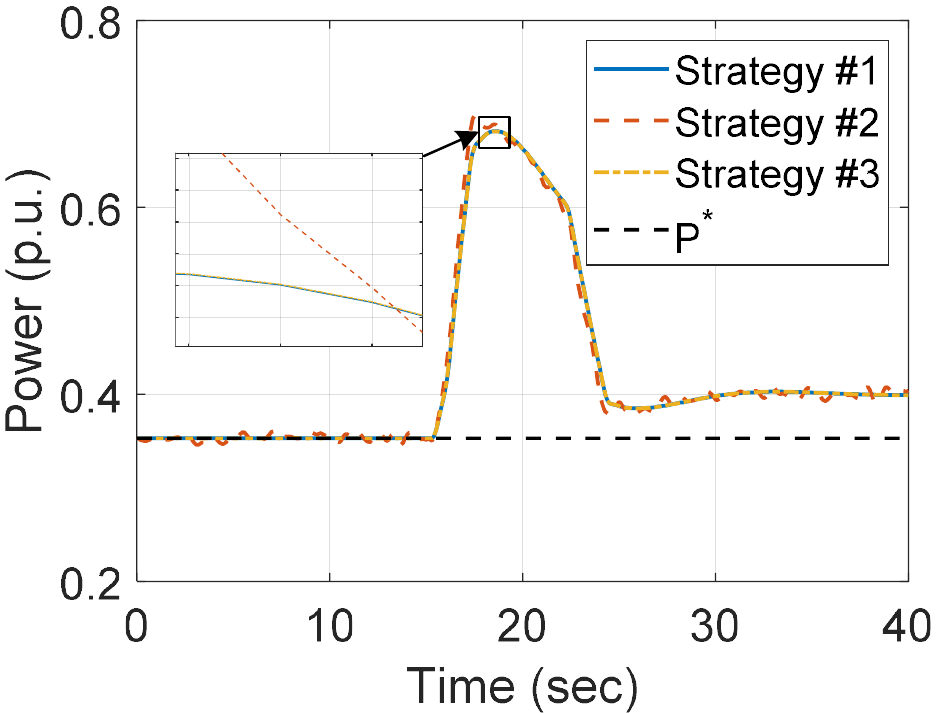}
  \caption{lower bound uncertainties}
  \label{fig:frob_robust_ucp_lower}
  \end{subfigure}
  \caption{FROB robustness considering uncertain system parameters.}
  \label{fig:frob_robust_ucp}
\end{figure}

The HPP might have several parameters that are constants but uncertain within a range. When actual parameters deviate from the nominal values used in the FROB design, the performance of FROB should be investigated. For example, control parameters of an asset converter vary around a nominal value that is used to construct the nominal WPP model for the FROB. Here, uncertainties in parameters are considered as:
\begin{equation}
\begin{split}
    \delta1 &= T_{cc\_MSC} \in [0.001, 0.1], \delta2 = K_{p\_pc\_MSC} \in [0.05, 0.15], \\
    \delta3 &= K_{i\_pc\_MSC} \in [5, 15], \delta4 = T_{cc\_GSC} \in [0.001, 0.1], \\
    \delta5 &= K_{p\_vc\_GSC} \in [7.5, 22.5], \delta6 = K_{i\_vc\_GSC} \in [75, 225].
\end{split}
\end{equation}
where $T_{cc\_MSC}$ is time constant of current control of machine-side converter, $K_{p\_pc\_MSC}$ and $K_{i\_pc\_MSC}$ are proportional and integral gains of power control of machine-side converter, $T_{cc\_GSC}$ is time constant of current control of grid-side converter, $K_{p\_vc\_GSC}$ and $K_{i\_vc\_GSC}$ are proportional and integral gains of DC-link voltage control of grid-side converter. Fig. \ref{fig:frob_robust_ucp_upper} and \ref{fig:frob_robust_ucp_lower} shows frequency responses for upper and lower bound uncertain systems. Strategy \#3 always provides frequency response identical to the ideal response (Strategy \#1). On the other hand, the deviation of system parameters causes the failure of estimating frequency responses from asset controllers using Strategy \#2. Thus, FROB shows robustness over uncertain systems.

\subsubsection{Impact of unknown FC malfunction}
\begin{figure}[t]
  \centering
  \includegraphics[scale = 0.85]{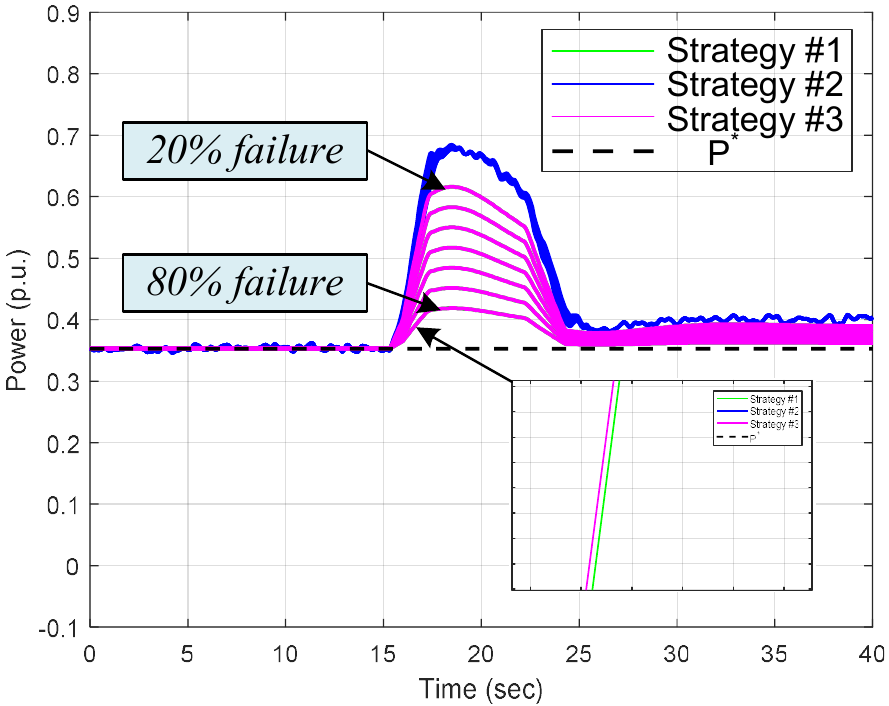}
  \caption{FROB robustness considering unknown FC malfunction.}
  \label{fig:frob_robust_ufcm}
\end{figure}

It happens when FC at some assets fail to respond to frequency dips due to device malfunction. When the failure is unknown to plant controller, whether frequency response is properly coordinated across control hierarchy needs to be investigated. Fig. \ref{fig:frob_robust_ufcm} shows that frequency response from Strategy \#3 adaptively coincides with that from Strategy \#1. It implies that FROB is robust in estimating the total frequency response, regardless of what percentage of assets experience FC failure. Meanwhile, frequency response from Strategy \#2 indicates overcompensation from plant controller, because the total frequency response from all the assets is overestimated with plant controller not being notified of those malfunctioning units.

\subsubsection{Impact of time-varying communication delay}
\begin{figure}[t]
  \begin{subfigure}{0.24\textwidth}
  \includegraphics[width = \textwidth]{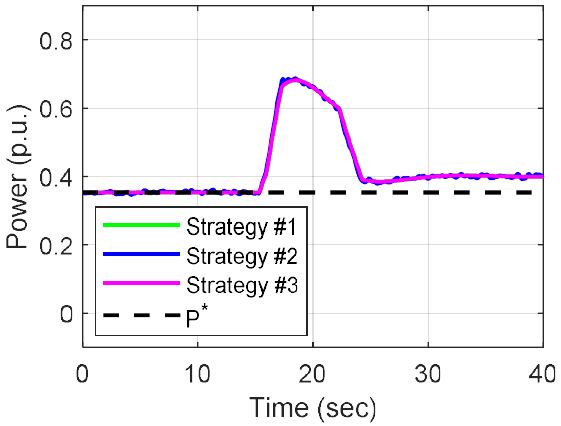}
  \caption{$T_{cd}$ = 0.0 s}
  \label{fig:frob_robust_vtd1}
  \end{subfigure}
  \hfill
  \begin{subfigure}{0.24\textwidth}
  \includegraphics[width = \textwidth]{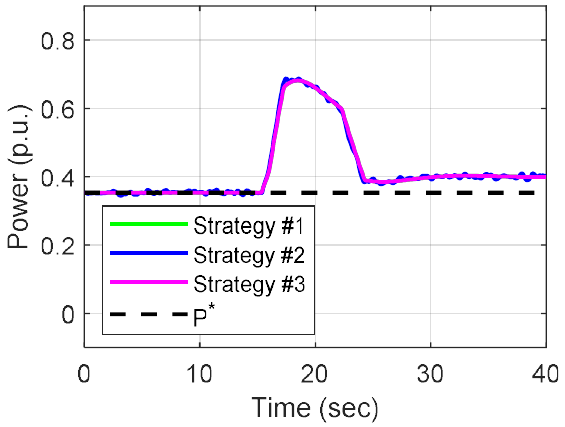}
  \caption{$T_{cd}$ = 0.1 s}
  \label{fig:frob_robust_vtd2}
  \end{subfigure}
  \medskip
  \begin{subfigure}{0.24\textwidth}
  \includegraphics[width = \textwidth]{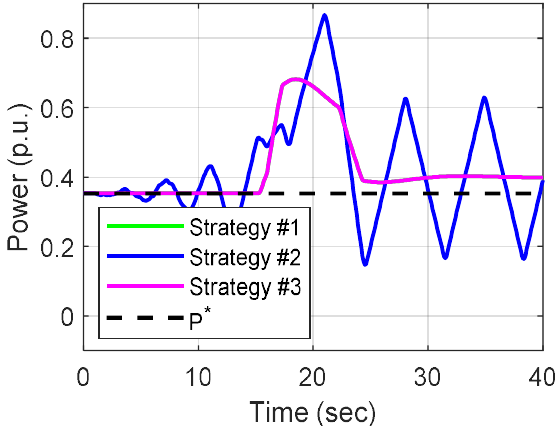}
  \caption{$T_{cd}$ = 1.0 s}
  \label{fig:frob_robust_vtd3}
  \end{subfigure}
\hfill
  \begin{subfigure}{0.24\textwidth}
  \includegraphics[width = \textwidth]{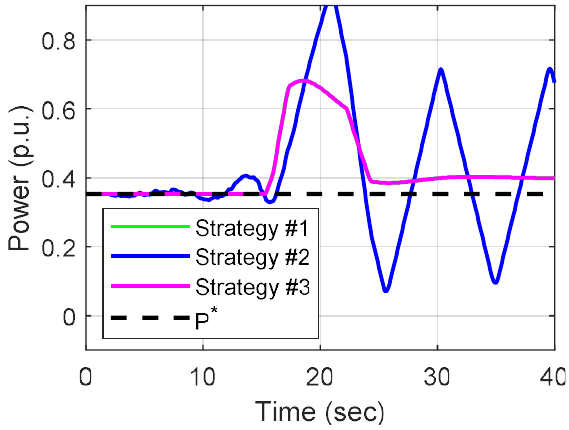}
  \caption{$T_{cd}$ = 2.0 s}
  \label{fig:frob_robust_vtd4}
  \end{subfigure}
  \caption{FROB robustness considering time-varying communication delay.}
  \label{fig:frob_robust_vtd}
\end{figure}

The exact value of communication delay between asset controllers and plant controllers is uncertain, and the interval $0 \leq T_{cd} \leq 2 s$ is used. Fig. \ref{fig:frob_robust_vtd} shows frequency response considering $T_{cd}$ equal to 0.0 sec, 0.1 sec, 1.0 sec and 2.0 sec. While Strategy \#2 experiences instability issues with large communication delay, there is no stability issue observed from Strategy \#3, which in all cases provides the identical frequency response to that of Strategy \#1. Although tuning control gains could be an option of resolving instability for Strategy \#2, it could still be challenging to obtain a set of control gains accommodating time-varying communication delays. With the FROB integrated into plant controllers, control gains can be kept the same as those designed for cases without communication latency. This characteristic shows the superior performance of the FROB regarding robustness against time-varying communication delay.

\section{Conclusion}

This paper proposes an innovative hierarchical FC approach that enables HPPs to provide FCSs. The proposed approach coordinates not only among multiple technology power plants but also across control hierarchy. Fast FCSs like FFR and FCR are implemented at asset level while slow FCS such as FRR is implmented at plant level and HPP level. A novel observer called FROB, which is inspired by frequency-domain DOBs, is developed to estimate the total frequency responses activated by asset controllers, for the purpose of avoiding control counteraction. In this way, the proposed approach accommodates state-of-the-art fast FC designed at asset level, coping with the increasingly demanding technical requirements, even when there are system uncertainties, such as uncertain control parameters, unknown malfunction or time-varying communication delays.


%



\section*{Acknowledgment}

The work is done as a part of Indo-Danish HYBRIDize project funded by Innovation Fund Denmark (IFD) with grant number 8127-00015B.

\ifCLASSOPTIONcaptionsoff
  \newpage
\fi



\bibliographystyle{IEEEtran}


%

%







\end{document}